# An Ultrahigh-Q Microresonator on 4H-silicon-carbide-on-insulator Platform for Multiple Harmonics, Cascaded Raman Lasing and Kerr Comb Generations

*Chengli Wang,* [1, 4, #] *Zhiwei Fang,* [2, #] *Ailun Yi,* [1, 4, #] *Bingcheng Yang,* [1, 4] *Zhe Wang,* [3, 4] *Liping Zhou,* [1, 4] *Chen Shen,* [1] *Yifan Zhu,* [1, 4] *Yuan Zhou,* [3, 4] *Rui Bao,* [2] *Zhongxu Li,* [1, 4] *Yang Chen,* [1, 4] *Kai Huang,* [1] *Jiaxiang Zhang,* [1, 4, a)] *Ya Cheng,* [2, 3, b)] *and Xin Ou,* [1, 4, c)]

[1] State Key Laboratory of Functional Materials for Informatics, Shanghai Institute of Microsystem and Information Technology, Chinese Academy of Sciences, Shanghai, 200050, China

[2] The Extreme Optoelectromechanics Laboratory (XXL), School of Physics and Electronic Science, East China Normal University, Shanghai 200241, China

[3] State Key Laboratory of High Field Laser Physics and CAS Center for Excellence in Ultra-intense Laser Science, Shanghai Institute of Optics and Fine Mechanics, Chinese Academy of Sciences, Shanghai 201800, China

[4] The Center of Materials Science and Optoelectronics Engineering, University of Chinese Academy of Sciences, Beijing 100049, China

# These authors contribute equally

Corresponding author email: a) jiaxiang.zhang@mail.sim.ac.cn; b) ya.cheng@siom.ac.cn; c) ouxin@mail.sim.ac.cn

(Date: May 6, 2021)

**Abstract**

The realization of ultrahigh quality (Q) resonators regardless of the underpinning material platforms has been a ceaseless pursuit, because the high Q resonators provide an extreme environment of storage of light to enable observations of many unconventional nonlinear optical phenomenon with high efficiencies. Here, we demonstrate an ultra-high Q factor ($7.1 \times 10^6$) microresonator on the 4H-silicon-carbide-on-insulator (4H-SiCOI) platform in which both $\chi^{(2)}$ and $\chi^{(3)}$ nonlinear processes of high efficiencies have been generated. Broadband frequency conversions, including second-, third-, fourth- harmonic generation were observed. Cascaded Raman lasing was demonstrated in the SiC microresonator for the first time to the best of our knowledge. Broadband Kerr frequency combs covering from 1300 to 1700 nm were achieved using a dispersion-engineered SiC microresonator. Our demonstration is a significant milestone in the development of SiC photonic integrated devices.

**Introduction**

High quality (Q) factor optical microresonators, with the capability of significantly enhancing light-matter interaction, have attracted strong interest in photonics community[1]. The novel photonic devices are highly in demand for both basic and applied research, such as cavity quantum electrodynamics[2], highly sensitive sensor[3], nonlinear devices or filter elements[4] for optical telecommunication systems, in which the high Q factors are crucial for achieving high spectral resolution and sensitivity as well as strong nonlinear light-matter interaction. For highly functional optical microresonators, the important requirements of the material platforms are ultralow optical loss, wide transparent window, high index contrast, high nonlinearities, and industry compatible fabrication processes. In the past few years, we have witnessed the great success of on-chip microresonators on various photonic platforms such as Si[4], $Si_3N_4$[5], GaAs[6] and $LiNbO_3$[7-9], etc.

Recently, silicon carbide has generated significant attention for its superior material properties to match all the essential requirements. As a mature wide bandgap material, SiC has a wide bandgap (3.26 eV for 4H polytypes), a high refractive index (2.6 at 1550 nm) and a wide transparent window (0.37 – 5.6 μm)[10], which can avoid multiple photon absorption that bothers the Si photonics. SiC is a CMOS-compatible semiconductor material, thus holds promise for realizing the monolithic integration of electronics and photonics with low fabrication costs via CMOS foundry[11], giving rise to more competitiveness than LiNbO$_3$ photonics. The non-centrosymmetric crystal structures of SiC grants both second-order (30 pm/V) and third-order (on the order $10^{-18}$ m$^2$/W) nonlinear effects[12], this enables access to efficient wideband frequency conversion and on-chip generation of non-classical light states. Moreover, unlike Si$_3$N$_4$ and Si, silicon carbide exhibits the Pockels effects and thus can be used for lossless, ultrafast and wide-bandwidth data transmission[13] which is unachievable in Si$_3$N$_4$ and Si photonics. In addition to the above advantages, the combination with its optically-addressable spin qubits[14], high breakdown voltage ($3 \times 10^{-6}$ V/cm), high thermal conductivity (4.9 W cm$^{-1}$ K$^{-1}$), and high optical damage threshold (80 GW/cm$^2$) further makes the SiC platform a unique and ideal candidate for realizing monolithic integration of electronics, quantum and nonlinear photonics[15,16].

SiC photonics has been developed for over a decade[17-23], one of the major obstacles for the practical application is the difficulty of fabricating ultralow optical loss SiC thin films on the wafer-scale. 4H-silicon-carbide-on-insulator (4H-SiCOI) formed by ion-cutting technique has been optimized[24], however, the material absorption generated by the ion-implantation-induced defects was considered as the main loss source[22] and the Q factor is limited to below $10^5$. Up to present, it is unclear whether the SiC thin films prepared by ion-cutting technique can be recovered to its pristine quality after the post thermal treatment. A different approach based on thin-film epitaxy techniques enabled microresonators[18,19,25] with Qs up to $2.5 \times 10^5$, which is still likely limited by material absorption[25]. Very recently, SiC thin films prepare by thinning of bulk wafer was demonstrated to obtain Qs about 1 million[21], which represents a vital and significant progress toward the high Q SiC photonics platform.

Here, we demonstrate an ultra-low loss 4H-SiCOI platform with a Q factor of record-high $7.1 \times 10^6$. The damage-free 4H-SiCOI photonics platform was prepared by wafer-bonding and thinning techniques. The ultrahigh-Q resonators were used to demonstrate various nonlinear processes including generation of multiple harmonics up to the fourth order, cascaded Raman lasing, and Kerr frequency comb. Broadband frequency conversions, including second-, third-, fourth- harmonic generation (SHG, THG, FHG) were observed. Cascaded Raman lasing with Raman shift of 204.03 cm$^{-1}$ was demonstrated in SiC microresonators for the first time. The Raman effect can be well controlled to enable a broadband Kerr frequency combs by tuning the pump wavelength. Using a dispersion-engineered SiC microresonator, Kerr frequency combs covering from 1300 to 1700 nm were achieved at a low input power of 13 mW.

**Material and method**

The ultra-high Q microresonators were fabricated on a pristine 4H-SiCOI wafer. The fabrication process of the wafer-scale 4H-SiCOI is schematically illustrated in Fig. 1(a). The 4-inch high-purity semi-insulating 4H-SiC wafer and the thermally oxidized Si (100) substrate were directly bonded at room temperature to form bulk-SiC-SiO$_2$-Si structure. Plasma surface activation was used in this process. In order to enhance the bonding strength, the bonded wafer was annealed at 600 °C in N$_2$ atmosphere for 8 hours. Then, the bonded wafer was processed by mechanical grinding to thin the SiC layer to a thickness of sub 10 μm. Fig.1(b) shows the image of the 4H-SiCOI substrate after grinding process. Over 95% of SiC thin film remained intact after this process. As the thickness measurement shown in Fig. 1(c), the fraction of uniform area within the thickness range of 2-4 μm beyond 60%. Finally, the wafer was cut into 10 mm × 12 mm chips, and each chip was further thinned to the predesignated thickness by inductively-coupled-plasma (ICP) reactive-ion-etching (RIE) in SF$_2$/O$_2$ plasma and chemo-mechanical polish (CMP). Fig. 1(d) shows a photograph of a 10 mm × 12 mm 4H-SiCOI chip with a thickness of 800 nm ± 80 nm, which illustrate an improvement comparing to previous results[15,23]. The uniformity of this area is sufficient to support an on-chip, compact photonic integrated circuit with rich functionalities[26] in the fields of telecommunication, nonlinear optics and quantum photonics. It should be noted that the thickness fluctuation of the SiC layer can be further reduced using foundry solutions, e.g., wafer trimming.

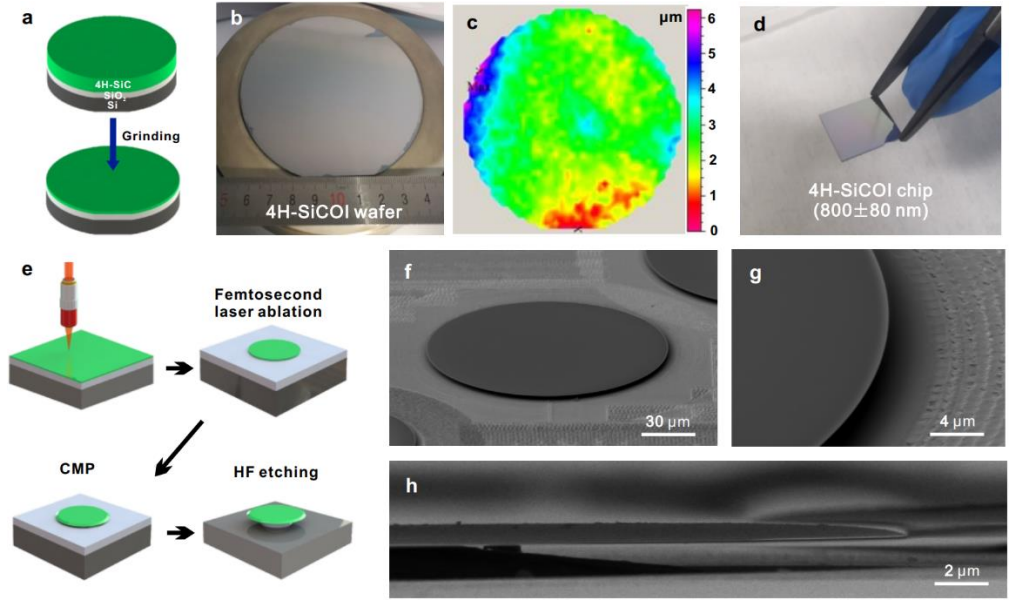

Fig. 1. (a) Fabrication process of pristine 4H-SiCOI material platform. (b) Photograph of 4-inch wafer-scale 4H-SiCOI substrate fabricated using bonding and thinning method. (c) The total thickness variation of 4H-SiCOI substrate. (d) Image of a 4H-SiCOI chip. (e) Flowchart of fabricating a SiC microdisk resonator. (f) A scanning electron micrograph (SEM) of the fabricated microdisk resonator. (g) Zoom-in SEM image of the sidewall of the resonator. (h) Side view SEM image of the fabricated resonator with parabolic shaped upper surface.

To investigate the optical quality of the prepared 4H-SiCOI, we fabricated microdisk resonators using femtosecond laser-assisted CMP method, which have been demonstrated in realizing ultrahigh Q LiNbO$_3$ resonators[27,28]. As schematically illustrated in Fig. 1(e), the prepared 4H-SiCOI structure is a layer of SiC (800 nm) on top of a buried silicon oxide (2 μm) layer on silicon substrate. To pattern the resonator, femtosecond laser micromachining was employed. This method has some unique characteristics including non-thermal ablation, high spatial resolution combined with decent material removal rates, as well as flexibility in generating arbitrary patterns in the mask-less direct write fashion. The femtosecond laser beam was focused into a ∼1 μm diameter focal spot using an objective lens (100×/NA 0.7), and the micromachining was carried out at a scan speed of 10 mm/s of the focused laser spot. It is noteworthy that femtosecond laser ablation generally leaves behind a surface roughness on the order of ∼100 nm, which should be eliminated for fabricating high-Q microresonators. The chemo-mechanical polishing (CMP) process was performed to smooth the top surface and sidewall of the 4H-SiC microdisk using a wafer lapping polishing machines. The CMP process allows to achieve an extremely low surface roughness of 0.15 nm at the edge of 4H-SiC microdisk, which is vital for achieving ultrahigh Q factors. Lastly, the suspended microdisk was formed by undercutting the silica layer into the pedestal in a diluted hydrofluoringe acid solution (10%). Fig. 1 (f) shows the scanning electron micrograph (SEM) images of a fabricated SiC microresonator with a diameter of 160 µm. The close-up view of the edge of the resonator is shown in Fig.1 (g), which reveals the achieved ultrasmooth surface and sidewall. The side-wall (upper surface) of the fabricate SiC microdisk has a parabolic shape.

The Q factor and nonlinear optical properties of the fabricated 4H-SiC microdisk resonators were examined with the measurement setup as shown in Fig. 2(a). Here, a C-band continuous-wave tunable laser (DLC CTL 1550, TOPTICA Photonics Inc.) was used as both the signal (for measuring Q) and pump (for exciting various nonlinear processes) source. The fine tuning of the laser was controlled by an arbitrary function generator (AFG3052C Tektronix Inc.). The polarization state of the tunable laser was adjusted by a fiber polarization controller (FPC562, Thorlabs Inc.). The tunable laser was amplified by an erbium-doped fiber amplifier (KY-EDFA-HP-37-D-FA, Beijing Keyang Optoelectronic Technology Co., Ltd.). A tapered fiber with a waist of 1 μm was used to evanescently couple the light into and out of the fabricated 4H-SiC microdisk. A photodetector (New focus 1811, Newport Inc.) was used to record the signal from the tapered fiber and to convert the optical signal to electrical signal. The electrical signal was further sent to an oscilloscope (MDO3104 Tektronix Inc.) for the Q factor measurement of the 4H-SiC microdisk resonator. To characterize the nonlinear optical properties of the fabricated 4H-SiC microdisk resonators, 90% of the output beam from the coupling fiber was directed to an optical spectrum analyzer (OSA: AQ6370D, YOKOGAWA Inc.) for infrared spectral analysis using a fiber beam splitter, whereas the remaining 10% of the output beam was routed to an ultraviolet-visible spectrometer (NOVA, Shanghai Ideaoptics Corp., Ltd) for ultraviolet-visible spectral analysis.

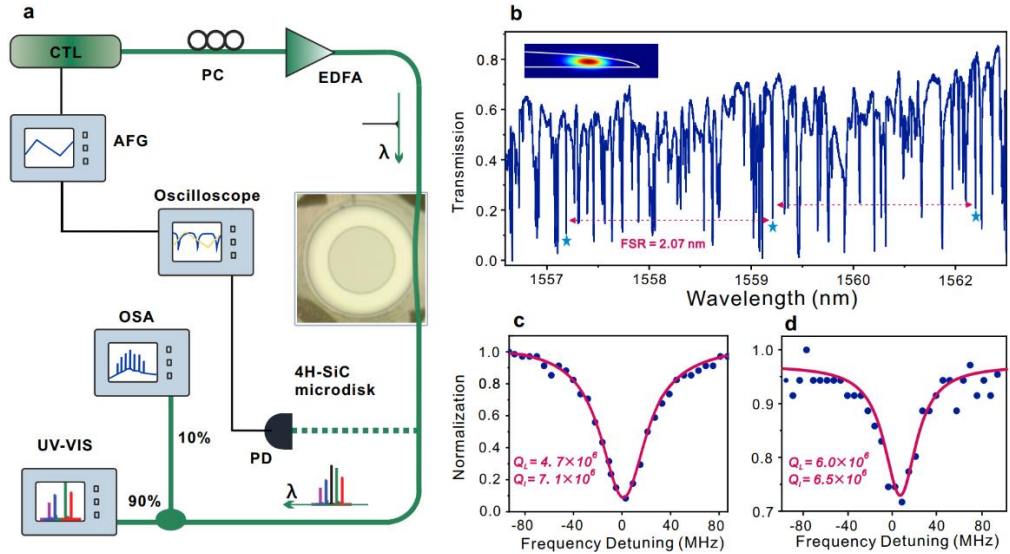

Fig. 2. (a) Diagram of the measurement setup for measurement of nonlinear optical processes in the microresonator. (b) Transmission spectrum of the SiC microresonator. Inset show the simulated mode profile of the fundamental TE mode. (c-d) High-resolution scan of the resonant mode. The Lorentz fitting reveals a Q-factor up to $7.1 \times 10^6$. (AFG: Arbitray function generator; CTL: Continuous-wave tunable laser; PC: Polarization controller; EDFA: Erbium-Doped Fiber Amplifier; UV-VIS: ultraviolet-visible spectrometer; OSA: Optical spectrum analyzer; PD: Photodetector)

Fig. 2(b) shows a typical transmission spectrum with multiple high-Q optical mode. A mode family labeled by star share a similar free-spectra-range (FSR) of 2.07 nm. By comparing the results calculated using finite-element simulation, we identified the labeled mode family to fundamental TE mode. The inset image shows its simulated mode profile. The coupling conditions of the optical mode can be adjusted by tuning the coupling positive finely. For example, the fundamental TE mode is nearly critically coupled is this case. This mode was chosen for the measurement of the Q factor by fitting with a Lorentz function as shown in Fig. 2(c). A load Q of $4.7 \times 10^6$ is calculated from the fitting curve, corresponding an intrinsic Q of $7.1 \times 10^6$. High resolution scan also found many resonant peaks whose full width at half maximum are narrower than the fundamental mode as shown in Fig. 2(a), the load Q and intrinsic Q are estimated to be $6.0 \times 10^6$ and $6.5 \times 10^6$, respectively. To the best of our knowledge, the Q factor is the highest among the demonstrated SiC microresonators so far[15,19,22,25].

**SHG, THG, and FHG generation**

Frequency conversions in the ultra-high Q SiC microresonators were investigated which occurred at the pump powers lower than that required by other nonlinear processes such as Raman lasing and comb generation as will be discussed later. As a proof of concept, the microresonators were not intentionally designed for satisfying the phase matching condition. When the pump laser wavelength was tuned between 1530 nm and 1570 nm, and the in-coupled power was set at 10 mW, strong emissions of various colors in the visible spectrum including red, orange, yellow, green, and purple light appear in the microdisk, which can be clearly capture by the CCD camera (see Visualization S1). The bright emissions can even be spotted by naked eye (see Visualization S2). The emitted spectrum was recorded by an OSA and a spectrometer through the tapered fiber when the pump wavelength was tuned around 1552.6 nm. In the visible spectrum shown in Fig. 3 (b-c), the emission peaks at 776.3 nm, 517.5 nm and 388.0 nm can be attributed to the second, third and fourth harmonic generation processes, respectively, since their wavelengths are exactly 1/2, 1/3 and 1/4 of the pump wavelength. The inset images in Fig.3 (b-d) show the red, green and purple light from the micro resonator captured by a visible CCD camera. Note that the ultralow photon counts of the recorded FHG is due to the low collection rate in our current experiment setup. Actually, the purple light from FHG can be clearly capture by CCD camera as shown by the inset images in Fig.3 (d), which reveals that the SiC microresonator exhibits the strong capability of FHG. The third and fourth harmonic generations observed in the current work are, to the best of our knowledge, the reported for the first time among the on-chip SiC photonics devices.

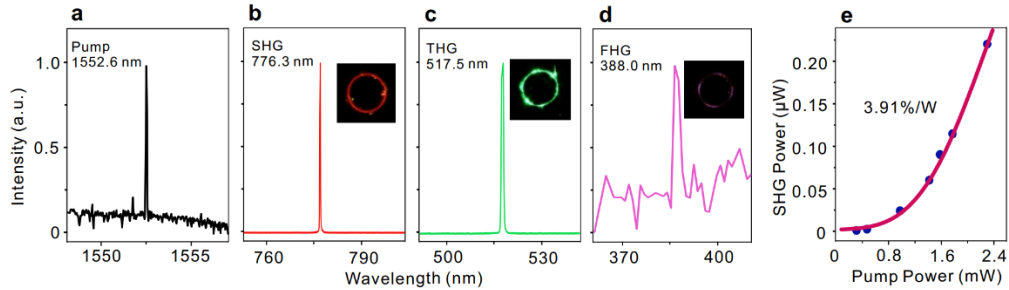

Fig. 3. Spectra of the pump light (a), the second-harmonic wave (b), the third-harmonic (c), and the fourth-harmonic wave (d). Inset: Top-view optical micrograph of the SHG, THG and FHG from the microresonator. (e) The filled circles show the dependence of the SHG power on the fundamental power. The redline is the fitting curve for the SHG data with a normalized conversion efficiency of 3.91%/W.

We also investigated the dependence of the SHG output power on the fundamental wave input power. The SHG power was measured by the OSA through the tapered fiber. As seen in the Fig. 3(e), the SHG intensity increases with input power. A fit to the data shows that the SHG intensity is proportional to the square of the input power, as expected for the second-order nonlinear process[29]. The solid line in Fig. 3(e) is the fitting curve following the assumed second-order nonlinearity relationship $P_{SHG} = \eta_{SHG} P_{Fundamental}^2$, where $\eta_{SHG}$ is the normalized SHG conversion efficiency. The fitting reveals that the $\eta_{SHG}$ is about 3.91 %/W. It should be noted that this value is greatly limited by the fiber taper collection efficiency for the SHG wavelength. Further improvements, such as collecting the visible signal by spatial collection through objective lens[30] or a butt/grating coupler[31], and designing the appropriate phase matching conditions, would make it possible to increase the conversion efficiency by at least two more orders of magnitude[32].

**Characterization of Raman Lasing**

For Raman lasing measurement, an amplified continuous-wave (CW) pump laser tuned to around 1552.2 nm was injected into the microresonator via the tapered fiber. The power of the pump light was gradually increased, and the laser wavelength was slowly tuned to the high-Q resonance mode at the same time. As the pump power reached above 9 mW, Raman lasing at 1603.3 nm was observed in the measured optical spectrum [Fig. 4(a)]. The Stokes lines show the frequency shifts of 204.03 cm$^{-1}$, corresponding to the phonon branches of $E_2$(TA) in the single-crystalline SiC[33]. When the pump power was gradually increased further, the first-order Raman comb and the second-order Raman lasing appeared successively. Interestingly, these two processes cannot exist at the same time, which indicates the competition between the Raman comb and cascaded Raman lasing.

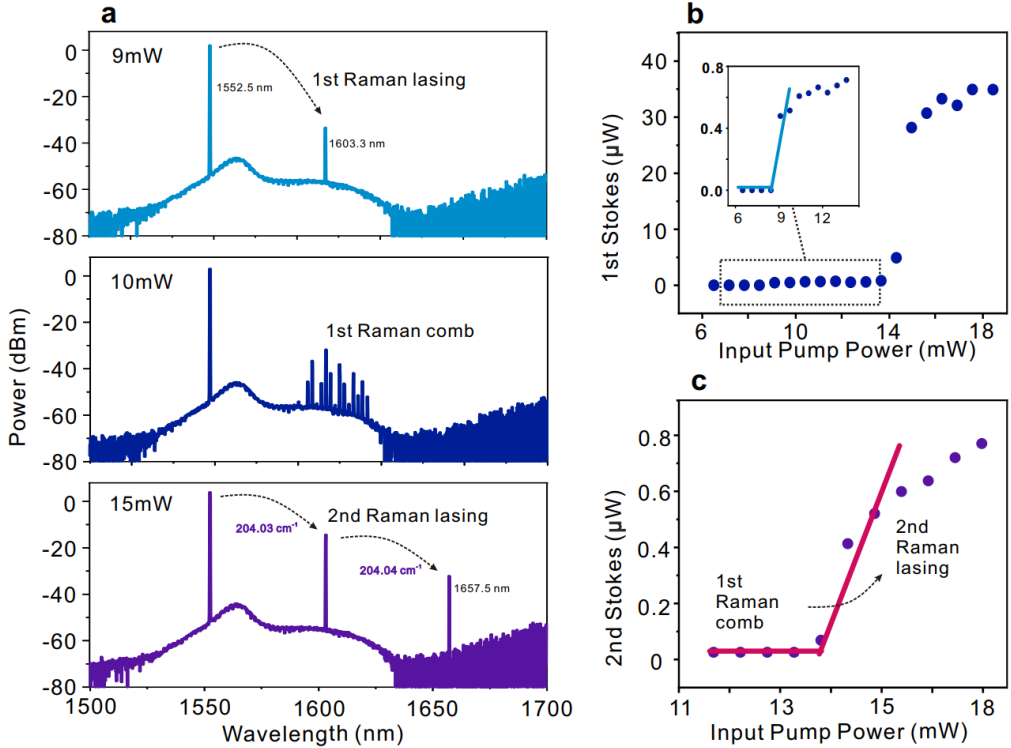

Fig. 4. Observation of cascaded Raman lasing and threshold measurement. (a) First-order Raman lasing at 1603.3 nm with a 204.3 cm-1 shift from the pump (top). First-order Raman comb generation (middle). Second-order Raman lasing generation (bottom). (b) The first-order SiC Raman laser output power as a function of the input pump power. The inset is a zoom-in of the data in the dotted box.

To illustrate this process, Fig. 4 (b) and (c) shows the measured output first-order and second-order Raman lasing as a function of the input pump power. Fig. 4(b) indicates that the first-order Raman lasing can be generated at two pump thresholds with the increasing pump power. The first threshold pump power is 10 mW, and the second one is 15 mW. Between the two threshold pump powers, a mini-comb is initiated around the first-order Raman lasing, and the output power of the first-order Raman lasing continues to slowly increase with increasing the pump power. As the pump power reaches about 14 mW, corresponding to the generated first-order Raman laser about 5 µW, the mini-comb disappears and the second-order Raman lasing action takes effect. The first-order laser also shows a large gain after this point, which may be due to the power transmission of the mini-comb back to first-order lasing. The output power of the second-order laser continues to increase by further enhancing the pump, while the first-order output power tends to be saturate at around 35 μW.

In the presence of Raman comb competition, the threshold pump power for the first-order Raman laser of the current SiC microresonator is measured to be 10 mW, whilst that for the cascaded second-order lasing in the same microresonator is 14 mW. Cascaded Raman lasing has been previously reported in optical fiber and other microresonators[34,35]. As a Raman-active media with the wide transparent window[10], this is the first demonstration of simulated Raman lasing and its cascaded process in the SiC photonic structures. Combining the unique material properties of SiC, such as high thermal conductivity and high optical damage threshold, the realization of the cascaded SiC Raman laser will offer a new opportunity to extend the spectral coverage of traditional laser light sources.

**Kerr comb generation**

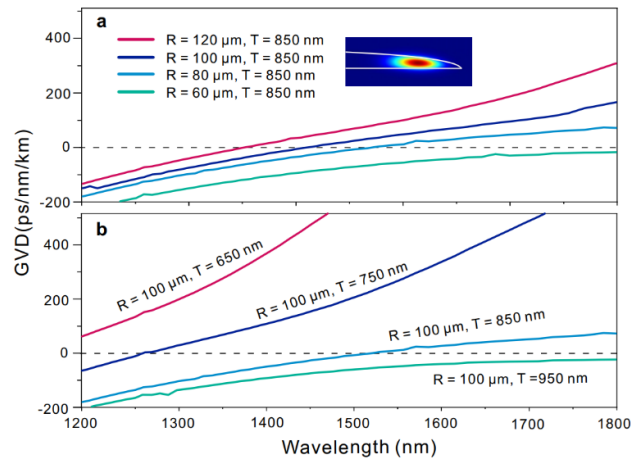

Fig. 5 (a) Dispersion calculation for the fundamental TM mode of the microresonators with a thickness of 850 nm and radius of 60, 80, 100, 120 μm. (b) Dispersion calculation for the fundamental TM mode of the microresonators with a radius of 100 μm and thickness of 650, 750, 850, 950 μm. Inset show the simulated mode profile of the fundamental TM mode.

In order to generate the broadband frequency comb, the record-high Q factor of the current SiC microresonator facilitates generation of the optical parametric oscillation (OPO) followed by the cascaded frequency conversion processes of high efficiency thank to the loss in-cavity losses[36]. Besides, an anomalous dispersion of the microresonator is also required to compensate for the nonlinear phase shift induce by self-phase modulation and cross-phase modulation[37]. To investigate the dispersion properties of the fabricated microresonator, we theoretically calculated the group-velocity dispersion (GVD) using a finite-element mode solver. The dispersion calculations include material anisotropy for TE and TM mode. Fig. 5 (a) shows the dispersion for the fundamental TM mode of the microresonators with different radius of 60, 80, 100, 120 μm, obviously, the dispersion curves increase with the increasing the resonator radius and can be tuned from normal dispersion to anomalous. Next, the thicknesses of 650, 750, 850, 950 nm were compared at the fixed radius of 100 μm as shown in Fig. 5(b). The calculation results show that the thinner microdisks provide greater anomalous dispersions. The SiC microresonators show the variable dispersion for the fundamental TM mode from normal to anomalous dispersion by controlling its thickness and radius. Note that the fundamental TE mode of the microresonator cannot be engineered to reach the anomalous dispersion regime.

Therefore, only the TM mode are excited in the following experiments for the Kerr comb generation. A microresonator with a radius of 100 μm and a thickness of 850 nm is chosen to meet the anomalous dispersion according to Fig. 5.

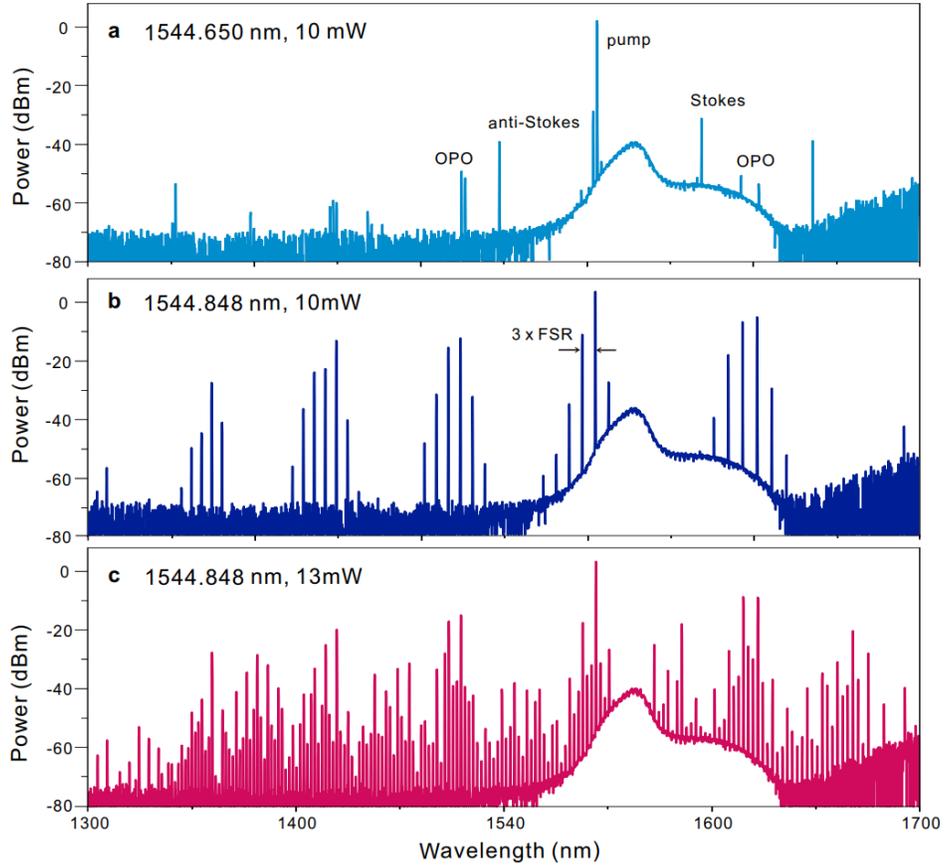

Fig. 6. Broadband Kerr frequency comb generations in a 4H-SiC microresonator with a radius of 100 um and thickness of 850 nm. (a) Measured OPO spectra generated with a launched pump power of 10 mW. (b) Hyper-OPO spectra generation when red-tuned the pump wavelength into resonance near 1544.848 nm. (c) Broadband Kerr frequency comb generations when a 13 mW pump was injected into the microresonator at 1544.848 nm.

The Kerr frequency comb is generated through the OPO process, which depends on the combination of parametric amplification and oscillation of nonlinear four wave mixing in microresonators. If a CW is injected and tuned into a high Q resonance, strong light field will be accumulated inside the microresonator. As the parametric gain exceeds the loss in a round-trip of the cavity, the accumulated power triggers OPO at the critical power threshold[38]. The OPO threshold power is measured to be around 10 mW as shown in Fig. 6(a) and (b). In Fig. 6(a), only the primary sidebands emerge when the pump wavelength at 1544.65 nm. The first-order Stokes, anti-Stokes and second-order Raman occur, which can be confirmed according to the corresponded wavelength shift of 204. 03 cm$^{-1}$. When the pump wavelength is gradually red-tuned into resonance near 1544.848 nm, the Raman-related signal vanishes but OPO oscillation persists, which indicate a power transfer between OPO and Raman oscillations. This power exchange is controllable and reversible by adjusting the pump frequency[39]. The OPO undergoes higher power gain until self-stabilized by thermal locking, which triggers more comb lines with a spectral spacing 3 times wider than the FSR. By further enhancing the pump power, the gaps between the primary sideband can be fulfilled by comb lines with spacing of one FSR (2.08 nm around 1550 nm). A broadband Kerr frequency comb spanning from 1300 to

1700 nm was measured as illustrated in Fig. 6(c). The shape of the observed combs indicates that the generated combs are modulation instability (MI) frequency combs[40]. For actual application, the next essential step is to access soliton formation. The thermo-optic coefficient of SiC ($4.21 \times 10^{-5}$ K$^{-1}$) is on the same order of magnitude compared with that of Si$_3$N$_4$ ($2.4 \times 10^{-5}$ K$^{-1}$) or silica ($0.8 \times 10^{-5}$ K$^{-1}$), and the Raman effects have been avoided in the current MI combs. The combination of the facts provides great prospect for soliton generation in SiC platform by using temporal scanning techniques, which has been widely demonstrated in other material platforms[40,41].

## Conclusion

During the preparation process of this manuscript, we noticed that Guidry *et. al.*[42] reported a SiC microresonator with a Q-factor of $5.6 \times 10^6$, which is realized using a similar thin flim fabrication process to ours. We, here, achieve a higher Q factor of $7.1 \times 10^6$. Using these devices, we demonstrate on-chip SHG with a conversion efficiency of 3.91 %/W without optimization of the phase matching. The third and fourth harmonic generations were observed for the first time in on-chip SiC photonics devices, which can be determined by the recorded spectrum and the bright color light emitted from the resonators. Cascaded Raman lasing was also demonstrated in SiC microresonators for the first time. The threshold pump powers were measured to be 10 and 14 mW for the first-order and cascaded second-order Raman lasing, respectively. Finally, low-threshold OPO and broadband (~ 400 nm) Kerr frequency combs were achieved using a dispersion-engineered SiC microresonator. We believe that the ultrahigh Q SiC photonics platform and its diverse nonlinear functionalities will pave the way to a wide range of quantum and classic applications based on 4H-SiCOI.

**Acknowledgments**

This work was supported by National Key R&D Program of China (2017YFE0131300, 2019YFA0705000), National Natural Science Foundation of China (No. U1732268, 61874128, 61851406, 11705262 ,11905282, 12004116, 12074400 and 11734009), Frontier Science Key Program of CAS (No. QYZDY-SSW-JSC032), Chinese-Austrian Cooperative R&D Project (No.GJHZ201950), Program of Shanghai Academic Research Leader (19XD1404600), Shanghai Sailing Program (No. 19YF1456200, 19YF1456400), K. C. Wong Education Foundation (GJTD-2019-11).

**Author contributions**

C.W., Z.F., A.Y., J.Z., Y.C., and X.O., conceived the experiments. C.W., A.Y. and X.O. developed the material fabrication process of 4H-Silicon-carbide-on-insulator. C.W., Z.F. and Y.C. developed the microdisk fabrication techniques. C.W., Z.F. and J.Z. carried out the device characterizations. C.W. and Z.F. performed the dispersion design, processed the experimental data, performed the analysis, drafted the manuscript. X.O. supervised the project. All the authors contributed to analysis of the data, discussions and the production of the manuscript.

**Data availability statement**

The data that support the plots within these paper and other findings of this study are available from the corresponding author upon reasonable request.

**Conflict of interest**

A.Y. and C.W. are involved in developing 4H-silicon-carbide-on-insulator technologies together with XOITEC corporation. All the authors declare that they have no conflict of interest.